\documentclass[%
    reprint,
    superscriptaddress,
    nofootinbib,
    amsmath,amssymb,
    aps,
    ]{revtex4-2}

\usepackage{hyperref}
\usepackage{cleveref,csquotes,placeins}
\usepackage[acronym]{glossaries}
\usepackage{lipsum}
\usepackage{graphicx}
\usepackage{dcolumn}
\usepackage{bm}
\usepackage{orcidlink}
\usepackage{booktabs}
\usepackage{mathtools,mathrsfs}
\usepackage{bbm}
\usepackage[shortlabels]{enumitem}

\DeclarePairedDelimiter\ket{\lvert}{\rangle}
\DeclarePairedDelimiterX\braket[2]{\langle}{\rangle}{#1\,\delimsize\vert\,\mathopen{}#2}

\usepackage{pgfplots}
\pgfplotsset{compat=newest}
\usepgfplotslibrary{external}
\usetikzlibrary{spy}
\usepgfplotslibrary{groupplots}
\pgfdeclarelayer{background}
\pgfdeclarelayer{foreground}
\pgfsetlayers{background,main,foreground}

\newacronym{tdse}{TDSE}{time-dependent Schrödinger equation}
\newacronym{tidse}{TIDSE}{time-independent Schrödinger equation}
\newacronym{qho}{QHO}{quantum harmonic oscillator}
\newacronym{dof}{DoF}{degrees of freedom}

\glsdisablehyper
\begin{document}

\rightline{YITP-26-40, RESCEU-11/26, IPMU26-0016\vspace{3mm}}

\title{Unitary Time Evolution and Vacuum for a Quantum Stable Ghost}%

\author{C\'edric~Deffayet\,\orcidlink{0000-0002-1907-5606}}
\email{cedric.deffayet@ens.fr}
\affiliation{Laboratoire de Physique de l'\'Ecole normale sup\'erieure, ENS, Universit\'e PSL, CNRS, Sorbonne Universit\'e, Universit\'e Paris Cit\'e, F-75005 Paris, France}

\author{Atabak Fathe Jalali\,\orcidlink{0009-0007-7717-3101}}
\email{jalali@fzu.cz}
\affiliation{CEICO--Central European Institute for Cosmology and Fundamental Physics, FZU--Institute of Physics of the Czech Academy of Sciences, Na Slovance 1999/2, 182 00 Prague 8, Czech Republic}
\affiliation{Institute of Theoretical Physics, Faculty of Mathematics and Physics, Charles University, V Hole\v{s}ovi\v{c}k\'ach 2, 180 00 Prague 8, Czech Republic}

\author{Aaron~Held\,\orcidlink{0000-0003-2701-9361}}
\email{aaron.held@phys.ens.fr}
\affiliation{Institut de Physique Théorique Philippe Meyer, Laboratoire de Physique de l’\'Ecole normale sup\'erieure (ENS), Universit\'e PSL, CNRS, Sorbonne Universit\'e, Universit\'e Paris Cité, F-75005 Paris, France}

\author{Shinji Mukohyama\,\orcidlink{0000-0002-9934-2785}} 
\email{shinji.mukohyama@yukawa.kyoto-u.ac.jp}
\affiliation{Center for Gravitational Physics and Quantum Information, Yukawa Institute for Theoretical Physics, Kyoto University, 606-8502, Kyoto, Japan}
\affiliation{Research Center for the Early Universe (RESCEU), Graduate School of Science, The University of Tokyo, Hongo 7-3-1, Bunkyo-ku, Tokyo 113-0033, Japan}
\affiliation{Kavli Institute for the Physics and Mathematics of the Universe (WPI), The University of Tokyo Institutes for Advanced Study, The University of Tokyo, Kashiwa, Chiba 277-8583, Japan}

\author{Alexander Vikman\,\orcidlink{ 0000-0003-3957-2068}} 
\email{vikman@fzu.cz}
\affiliation{CEICO--Central European Institute for Cosmology and Fundamental Physics, FZU--Institute of Physics of the Czech Academy of Sciences, Na Slovance 1999/2, 182 00 Prague 8, Czech Republic}


\begin{abstract}
We quantize a classically stable system of a harmonic oscillator polynomially coupled to a ghost with negative kinetic energy. We prove that due to an integral of motion with a positive discrete spectrum: i) the Hamiltonian has a pure point spectrum unbounded in both directions, ii) the evolution is manifestly unitary, iii) the vacuum is well-defined, iv) expectation values for squares of canonical variables are bounded. Numerical solutions of the Schr\"odinger equation confirm these results. We argue that the discrete spectrum of the integral of motion enforces stability for extended interactions.
\end{abstract}

\maketitle

{ \textbf{\textit{Introduction.}} }
It is a common lore that ghosts -- degrees of freedom with negative kinetic terms -- lead to fundamental problems with stability and unitarity, see e.g. \cite{Pais:1950za,Woodard:2006nt,Woodard:2015zca}. Ghost instabilities arise purely due to interactions, and QFT estimations imply a parametrically strong, if not instantaneous instability rate, see e.g. \cite{Cline:2003gs,Carroll:2003st}. However, ghosts often emerge in theories of gravity modified either in the UV, where higher derivatives are needed for renormalization \cite{Stelle:1976gc,Stelle:1977ry,Woodard:2009ns}, or in the IR, see e.g. \cite{Clifton:2011jh,Joyce:2014kja}.  The appearance of ghosts is a consequence of Ostrogradsky's theorem, see \cite{Ostrogradsky:1850fid,Woodard:2015zca}. 
Ghostly systems are of particular interest for cosmology in view of recent DESI results \cite{DESI:2025zgx,DESI:2025fii} and as means to avoid the origin-of-time cosmological singularity, for review see e.g. \cite{Brandenberger:2016vhg}. 
Recent work \cite{Deffayet:2021nnt,Deffayet:2023wdg} has clearly demonstrated the existence of such ghostly classical mechanical systems — with a finite number of interacting degrees of freedom — that admit not only local Lyapunov stability of the equilibrium points but also global stability.
Thus, the motion avoids runaways -- it is bounded to finite regions of phase space for all initial data and for all times. The stability proof from \cite{Deffayet:2021nnt,Deffayet:2023wdg} rests on the existence of an additional integral of motion that is not only \emph{positive definite}, but is also \emph{confining}\footnote{See e.g. \cite{Robert:2006nj,Kaparulin:2014vpa,Boulanger:2018tue,Smilga:2020elp,Damour:2021fva,Felski:2026grs} for related studies of theories with integrals of motion.} the motion in all directions in phase space, although the Hamiltonian is unbounded from above and from below. However, it seems that even classical field theories, devoid of any integrals of motion, may not reveal any instabilities for parametrically long times in numerical simulations~\cite{Deffayet:2025lnj,Held:2025fii}. Note, that such equations of motion are hyperbolic PDEs, which introduces a universal, wavelength-independent bound on any potential instability \cite{Deffayet:2025lnj}. Moreover, small perturbations can even be stable for arbitrary long times \cite{Held:2025fii}. 

In this Letter, we canonically quantize one of the mechanical models presented in \cite{Deffayet:2023wdg}. We demonstrate that a usual oscillator interacting with a ghost can be described by standard quantum mechanics with unitary evolution and with a stable unique vacuum, despite the Hamiltonian possessing a pure point spectrum unbounded from above and from below. Previous works on canonical quantization of ghost systems, involving standard unitarity, Hermitian operators and inner product, include \cite{Pais:1950za,Smilga:2005gb,Robert:2006nj,Ilhan:2013xe, Smilga:2017arl,Gross:2020tph,Smilga:2020elp,Fring:2025zha}. None of these works contain the results of this Letter for the interacting case. 

{ \textbf{\textit{Classical Model.}} }
Let us quantize the classical theory of an ordinary unit-mass oscillator $x$ coupled to a ghostly oscillator $y$ through an interaction potential $V_I(x,y)$. Thus, the $y$-oscillator has negative unit mass. We focus on the model presented in Section III.C of \cite{Deffayet:2023wdg}:
\begin{align}
    H =\, &\frac{p_x^2+x^2}{2}-\frac{p_y^2+\omega^2y^2}{2}+V_I(x,y)\,,\label{eq:system}\\
    V_I=\,&\frac{1}{2c}(1-\omega^2)(x^2-y^2)^2+\gamma\left[(x^2-y^2)^3+c(x^4-y^4)\right]\,,\notag
\end{align}
where the coupling constants $\gamma$ and $c$ are both \emph{positive}. The latter property ensures classical global stability -- boundedness of classical motion for all initial data, as we have analytically proven in \cite{Deffayet:2023wdg}. The decoupling limit corresponds to $\gamma\rightarrow 0$, $c\rightarrow\infty$, in such way that $c\gamma$ is either vanishing or finite. In the latter case the decoupled $x$ and the ghost $y$ are both stable quartic anharmonic oscillators. Note that we have simplified the notation in \cite{Deffayet:2023wdg} by replacing ${\omega_x\rightarrow1}$, ${\omega_y\rightarrow\omega}$, ${\tilde{c}\rightarrow c}$, and ${\mathcal{C}_{4}\rightarrow\gamma}$. Thus, time is measured in units of the angular frequency of the $x$-oscillator. 

Dynamical system \eqref{eq:system} possesses\footnote{To be precise, $\mathcal{H}_\uparrow=\frac{2}{c}J_{\text{LV}}$, where $J_{\text{LV}}$ is the integral of motion defined in equations (3.31), (3.32) and (3.70) in the journal version of \cite{Deffayet:2023wdg}.} an integral of motion 
\begin{equation}
{H}_{\uparrow}=p_x^2+p_y^2+\frac{2}{c}K^2+V_{\uparrow}\left(x,y\right)\,,\label{eq:classical_H+}
\end{equation}
where nontrivial kinetic interactions are given through the generator of boosts in $\left(x,y\right)$ space 
\begin{equation}
K=x\,p_{y}+y\,p_{x}\,,\label{eq:k}
\end{equation}
while the potential part is defined as
\begin{align}    V_{\uparrow}=\,&x^2+\omega^2y^2+\frac{1}{c}\left(1-\omega^{2}\right)\left(x^{2}-y^{2}\right)\left(x^{2}+y^{2}\right)\label{eq:V+}\\
    \notag&+2\gamma\left(\left(x^{2}-y^{2}\right)^{2}\left(x^{2}+y^{2}\right)+c\left(x^{4}+y^{4}\right)\right)\,.
\end{align}
It is straightforward to verify that the Poisson brackets with the total Hamiltonian vanish: $\left\{ H_{\uparrow},H\right\} =0$. 
If $H_{\uparrow}$ is interpreted as a Hamiltonian for some auxiliary motion, it describes two \emph{usual ghost-free} harmonic oscillators coupled through ${K^2}$ and ${V_{\uparrow}}$. The $x_{\uparrow}$-oscillator has mass $1/2$ and angular frequency $2$, while the $y_{\uparrow}$-oscillator has mass $1/2$ and angular frequency $2\omega$. 

It is convenient to introduce decomposition 
\begin{equation}\label{Split}
H=H_{\uparrow}-H_{\downarrow}\,,
\end{equation}
so that $H_{\downarrow}$ is another integral of motion given by
\begin{equation}
H_{\downarrow}=\frac{1}{2}p_{x}^{2}+\frac{3}{2}p_{y}^{2}+\frac{2}{c}K^2+V_{\downarrow}\left(x,y\right)\,,\label{eq:classical_H-}
\end{equation}
where the potential is given by 
\begin{align}
\notag
V_{\downarrow}=\,&\frac{1}{2}x^{2}+\frac{3}{2}\omega^{2}y^{2}+\frac{1}{2c}\left(1-\omega^{2}\right)\left(x^{2}-y^{2}\right)\left(x^{2}+3y^{2}\right)\\ &+\gamma\left(\left(x^{2}-y^{2}\right)^{2}\left(x^{2}+3y^{2}\right)+c\left(x^{4}+3y^{4}\right)\right)\,.\label{eq:V-}
\end{align}
Here there are again two coupled oscillators: the $x_{\downarrow}$-oscillator has a unit mass and a unit angular frequency, while the $y_{\downarrow}$-oscillator has mass $1/3$ and angular frequency $3\omega$. 
For ${x^{2}+y^{2}\rightarrow\infty}$, both potentials $V_{\uparrow\downarrow}$ grow without bound.
We will be interested in the choice of parameters  $\omega$, $c$, and $\gamma$ when both $V_{\uparrow\downarrow}$ have unique global minimum at the origin where they are vanishing, so that the system belongs to case III from Fig.~9 of  \cite{Deffayet:2023wdg}. 
A sufficient condition for this is 
\begin{equation}
2\gamma c^{2}\geq\left|1-\omega^{2}\right|\,,\label{condition_cg}
\end{equation}
which ensures that the quartic terms in both $V_{\uparrow\downarrow}$ are positive definite. In this case both $H_{\uparrow\downarrow}$ are positive definite,  
while by construction, ${\left\{ H_{\uparrow},H_{\downarrow}\right\} =0}$.

{ \textbf{\textit{Quantization.}} } 
We follow the standard postulates of canonical quantization. In the Schr\"odinger representation, the Hilbert space $\mathscr{H}$ can be taken to be ${L^2(\mathbb{R}^2)}$ -- the space of all square-integrable complex valued functions ${\Psi\left(x,y\right)}$, see e.g. \cite{Glimm:1987ylb} -- with \emph{the standard} $L^2$ inner product between any two states $\Psi,\Phi\in\mathscr{H}$:
\begin{equation}
\label{eq:standard_product}
\braket{\Psi}{\Phi} \equiv \int_{\mathbb{R}^2} dxdy\,\overline{\Psi}\left(x,y\right)\Phi\left(x,y\right)\,,
\end{equation}
where $\overline{\Psi}$ denotes the complex conjugate of $\Psi$. Note that numerous other papers devoted to quantization of ghostly systems, see e.g.~\cite{Bender:2007wu,Salvio:2015gsi,Strumia:2017dvt,Holdom:2024onr}, used different \emph{non-standard} inner products invalidating usual correspondence principle and semiclassical approximation. 

Then, the standard canonical commutation relations (with $\hbar=1$ used throughout the Letter)
\begin{align}\label{eq:commutation_relations}
    &\quad\quad\quad\quad[\hat{x},\hat{p}_x]=[\hat{y},\hat{p}_y]=i\,,\notag\\
    &[\hat{x},\hat{y}]=\left[\hat{x},\hat{p}_{y}\right]=\left[\hat{y},\hat{p}_{x}\right]=[\hat{p}_x,\hat{p}_y]=0\,,
\end{align}
are satisfied for ${\hat{x}=x}$, ${\hat{p}_x=-i\partial_x}$ and ${\hat{y}=y}$, ${\hat{p}_y=-i\partial_y}$. Crucially, there are no operator ordering ambiguities\footnote{In particular, as ${\hat{K}=-ix\partial_{y}-iy\partial_{x}=-i\partial_{y}x-i\partial_{x}y}$.} in promoting 
$H$, ${H_{\uparrow}}$, ${H_{\downarrow}}$ to operators $\hat{H}$, ${\hat{H}_{\uparrow}}$, ${\hat{H}_{\downarrow}}$. The standard time-dependent Schr\"odinger equation reads
\begin{equation}
\label{eq:Sch}
i\frac{d}{d t}\ket{\Psi(t)} =\hat{H}\ket{\Psi(t)} \,,
\end{equation}
where in the Schr\"odinger representation Hamiltonian \eqref{eq:system} is given by the symmetric \emph{hyperbolic} operator 
\begin{equation}\label{eq:quantized_system}
    \hat{H}=-\frac{1}{2}\left(\partial_{x}^{2}-\partial_{y}^{2}\right)+\frac{1}{2}\left(x^{2}-\omega^{2}y^{2}\right)+V_{I}\left(x,y\right)\,.
\end{equation}
On the other hand, 
\begin{equation}
\hat{H}_{\uparrow}=-\partial_{i}\left(M_{\uparrow}^{ik}\partial_{k}\right)+V_{\uparrow}\left(x,y\right)\,,
\end{equation}
where the indices $i,k$ take values in ${x,y}$ and where the symmetric matrix ${M_{\uparrow}^{ik}\left(x,y\right)}$ has components 
\begin{equation}
M_{\uparrow}^{xx}=1+\frac{2}{c}y^{2}\,,\quad M_{\uparrow}^{yy}=1+\frac{2}{c}x^{2}\,,\quad
M_{\uparrow}^{xy}=\frac{2}{c}xy\,,    
\end{equation}
so that it is positive definite, as ${\text{det }M_{\uparrow}=1+2\left(x^{2}+y^{2}\right)/c}$. A similar expression with a slightly different, positive definite, $M_{\downarrow}^{ik}$ holds for $\hat{H}_{\downarrow}$. Thus, both $\hat{H}_{\uparrow\downarrow}$  are \emph{uniformly elliptic operators of the second order} for any finite region in $(x,y)$ space. Following Theorem 12.33 and metatheorem 12.37 from \cite{Cycon:1987xw}
the operators $\hat{H}_{\uparrow\downarrow}$ are essentially self-adjoint. This, and the fact that ${\hat{H}}$ is also essentially self-adjoint, also follows from \cite{Reed:1975}, Theorem X.39, (Nelson's analytic vector theorem), see also Example 3, page 205. Thus, by the spectral theorem, the individual spectra of all these operators are entirely real. 
It is straightforward to check that $\hat{H}$, ${\hat{H}_\uparrow}$, and ${\hat{H}_\downarrow}$ remain in involution, 
\begin{equation}
\label{commute}
[\hat{H},\hat{H}_{\uparrow}]=[\hat{H},\hat{H}_{\downarrow}]=0\,,\quad[\hat{H}_{\uparrow},\hat{H}_{\downarrow}]=0 \,.
\end{equation}
Thus, any initial state ${\ket{\Psi(0)}}$ evolves to ${\ket{\Psi(t)}}~=~\hat{U}\left(t\right){\ket{\Psi(0)}}$ via unitary operator
\begin{equation}
\label{eq:back_forward}
\hat{U}\left(t\right)=e^{-it\hat{H}}=e^{it\hat{H}_{\downarrow}}e^{-it\hat{H}_{\uparrow}}=e^{-it\hat{H}_{\uparrow}}e^{it\hat{H}_{\downarrow}}\,.
\end{equation}
The unitary time evolution above can be interpreted as the motion of a system with Hamiltonian $\hat{H}_{\downarrow}$ backward in time followed by the motion of a system $\hat{H}_{\uparrow}$ forward in time (or vice versa). Thus, the evolution seen in Eq.~\eqref{eq:back_forward} is manifestly unitary, as each of this auxiliary motions is unitary, see Fig.~\ref{fig:sine-cosine} . Moreover, Eq.~\eqref{commute} implies that for any quantum state ${\left|\Psi\right\rangle}$ one has ${\frac{d}{dt}\left\langle \Psi\right|\hat{H}_{\uparrow}\left|\Psi\right\rangle =0}$ and similarly for $\hat{H}_{\downarrow}$. 

{ \textbf{\textit{Spectrum and Ground State.}} } One can notice that ${\hat{H}_{\uparrow}-\frac{2}{c}\hat{K}^{2}=-\Delta+V_{\uparrow}\left(x,y\right)}$ has the form of the usual Schr\"odinger operator with positive smooth potential growing at infinity. The latter is known to possess only a purely discrete positive spectrum, see Theorem XIII.16 \cite{Reed:1978}. Yet, from ${0\leq(\hat{H}_{\uparrow}-\frac{2}{c}\hat{K}^{2})\leq\hat{H}_{\uparrow}}$ and theorem 1.5.9 of \cite{Glimm:1987ylb}, see also \cite{Simon:1983jy}, we conclude that the spectrum of $\hat{H}_{\uparrow}$ is discrete and positive. Thus, eigenvalues $E_{n}^{\uparrow}>0$ can be sorted in ascending order by integer $n$  taking values in ${0, 1, 2, 3....}$. Furthermore, in each region $\Omega$ where ${\Psi\left(x,y\right)\neq0}$ one can use the polar decomposition  ${\Psi\left(x,y\right)=\rho\left(x,y\right)\exp\left( i\theta\left(x,y\right)\right)}$. Then  $\langle\hat{H}_{\uparrow}\rangle$ can be expressed in Dirichlet form and then decomposed in each region $\Omega$, so that 
\begin{align}
\notag
\left\langle \Psi\right|\hat{H}_{\uparrow}\left|\Psi\right\rangle =&\underset{\Omega}\sum\int_{\Omega} dxdy\big(M_{\uparrow}^{ik}\left(x,y\right)\partial_{i}\rho\,\partial_{k}\rho+V_{\uparrow}\left(x,y\right)\rho^{2}
\\
&\qquad\qquad+\rho^{2}M_{\uparrow}^{ik}\left(x,y\right)\partial_{i}\theta\,\partial_{k}\theta\big)\,.\label{eq:Dirichlet}
\end{align}
 The positive-definiteness of $M_{\uparrow}^{ik}\left(x,y\right)$ and $V_{\uparrow}\left(x,y\right)$ implies that 
\begin{equation}
\left\langle \Psi\right|\hat{H}_{\uparrow}\left|\Psi\right\rangle \geq\left\langle \rho\right|\hat{H}_{\uparrow}\left|\rho\right\rangle \geq 0\,\,,
\end{equation}
for an arbitrary state. Thus, the first Beurling-Deny criterion (Theorem XIII.50 from \cite{Reed:1978}) is satisfied and $e^{-\tau\hat{H}_{\uparrow}}$ is \emph{positivity preserving} for all $\tau>0$. Furthermore, as both $M_{\uparrow}^{ik}\left(x,y\right)$ and $V_{\uparrow}\left(x,y\right)$ are non-singular, all states $\Psi\left(x,y\right)$ are defined in a simply connected space. Hence, one can argue, analogously to Theorems XIII.47-48 \cite{Reed:1978}, that the ground state $|E_{0}^{\uparrow}\rangle$ of $\hat{H}_{\uparrow}$ is non-degenerate, and can be chosen to be positive, $\langle x,y|E_{0}^{\uparrow}\rangle>0$. All of the above applies to $\hat{H}_{\downarrow}$ as well. Since $\hat{H}_{\uparrow}$ and $\hat{H}_{\downarrow}$ commute, they have a shared and complete (orthogonal) system of eigenvectors, $\left|E_{n}^{\uparrow},E_{m}^{\downarrow}\right\rangle$. Due to \eqref{commute} and decomposition \eqref{Split} this basis corresponds to eigenvectors of the Hamiltonian 
 \begin{equation}
 \label{eq:E_spectrum}
\hat{H}\left|E_{n}^{\uparrow},E_{m}^{\downarrow}\right\rangle =\left(E_{n}^{\uparrow}-E_{m}^{\downarrow}\right)\left|E_{n}^{\uparrow},E_{m}^{\downarrow}\right\rangle\,,
 \end{equation}
 so that $\hat{H}$ does not have an essential spectrum, but only a pure point spectrum, $\mathcal{E}_{nm}=E_{n}^{\uparrow}-E_{m}^{\downarrow}$, unbounded from above and from below\footnote{We are aware of only one other model \cite{Smilga:2020elp} where the spectrum for interacting system with kinetic energies unbounded from below is claimed to be a pure point under standard canonical quantization. However, there the Hamiltonian is not quadratic in momenta and does not decouple to  oscillators.}. Yet, one can still define the {\emph{unique ground state}} or the \emph{vacuum} of the system. As $|E_{0}^{\uparrow}\rangle$ is nondegenerate and shared with eigenvectors of $\hat{H}_{\downarrow}$ it should correspond to a $|E_{0}^{\uparrow},E_{n}^{\downarrow}\rangle$. Then, due to $\langle x,y|E_{0}^{\uparrow}\rangle>0$ and $\langle x,y|E_{0}^{\downarrow}\rangle>0$, the product \eqref{eq:standard_product} gives $\langle E_{0}^{\uparrow},E_{n}^{\downarrow}|E_{0}^{\downarrow}\rangle>0$ which violates $\langle E_{n}^{\downarrow}|E_{0}^{\downarrow}\rangle=0$ for all $n\neq 0$. 
Thus, the \emph{ground states should coincide}, $|E_{0}^{\uparrow}\rangle=|E_{0}^{\downarrow}\rangle=|E_{0}^{\uparrow},E_{0}^{\downarrow}\rangle$. This argument is applicable for arbitrary coupling constants satisfying condition \eqref{condition_cg}. We denote this vacuum wavefunction as $\Psi_{0}\left(x,y\right)=\langle x,y|E_{0}^{\uparrow},E_{0}^{\downarrow}\rangle$. In particular, uniqueness of vacuum is obvious when one neglects all interactions\footnote{In that case the eigenvalues are not independent $E_{n}^{\downarrow}=f(E_{m}^{\uparrow})$.}. In that case,  $\Psi_{0}\left(x,y\right)=\Phi_0\left(x,y\right)$, where
\begin{equation}\label{eq:nonint_gs}
\Phi_0\left(x,y\right)=\left(\frac{\omega}{\pi^{2}}\right)^{\frac{1}{4}}\exp\left(-\frac{x^{2}+\omega y^{2}}{2}\right)\,,
\end{equation}
is a ground state for both, decoupled 2D harmonic oscillators $\hat{H}_{\uparrow\downarrow}$ and decoupled 1D oscillators $x$ and $y$. Splitting $\hat{H}$ into the difference of other two \emph{positive definite} integrals of motion given by a real linear combination of $\hat{H}_{\uparrow\downarrow}$ would result in an analogous construction, due to commutativity of these different integrals of motion with $\hat{H}_{\uparrow\downarrow}$ and their positive definiteness. 
Physically, the ground state $\Psi_{0}\left(x,y\right)$ still corresponds to the most stable state - to the minimum of the positive definite integrals of motion\footnote{In \cite{Gross:2020tph} it was found that ${\Phi_0\left(x,y\right)}$ is only metastable for other interactions.}. Moreover, as the interaction part of the potential ${V_{\uparrow}\left(x,y\right)-x^{2}-\omega^{2}y^{2}}$ in \eqref{condition_cg} is non-negative, it follows that for an arbitrary state 
\begin{align}
\notag
\langle \hat{H}_{\uparrow}\rangle =&\left\langle \hat{p}_{x}^{2}\right\rangle +\left\langle x^{2}\right\rangle +\left\langle \hat{p}_{y}^{2}\right\rangle +\omega^{2}\left\langle y^{2}\right\rangle +\frac{2}{c}\langle \hat{K}^{2}\rangle 
\\
&\qquad+
\langle \left(V_{\uparrow}-x^{2}-\omega^{2}y^{2}\right)\rangle=\text{const}>0\,,\label{eq:bound} 
\end{align}
resulting in the averages ${\left\langle \hat{p}_{x}^{2}\right\rangle}$, ${\left\langle \hat{p}_{y}^{2}\right\rangle}$, ${\left\langle x^{2}\right\rangle}$, ${\left\langle y^{2}\right\rangle}$ being all bounded (see Fig.~(\ref{fig:confinement})). 
Intuitively it is clear that they are minimal in the ground state ${\Psi_{0}\left(x,y\right)}$, when ${\langle \hat{H}_{\uparrow}\rangle}$ is minimal, and that they only grow with higher levels of $\hat{H}_{\uparrow}$ or $\hat{H}_{\downarrow}$. Drawing on minimal fluctuations, positivity, non-degeneracy, and on intuition from the decoupled case, we declare ${\Psi_{0}\left(x,y\right)}$ to be the ground state of the system. Clearly, ${\Psi_{0}\left(x,y\right)}$, does not correspond to a minimum of Hamiltonian, but to an eigenvalue ${\mathcal{E}_{0}\equiv\mathcal{E}_{00}=E_{0}^{\uparrow}-E_{0}^{\downarrow}=\frac{1}{2}\left(1-\omega\right)+\mathcal{O}\left(\gamma,c^{-1}\right)}$. 
Discrete spectrum of $\hat{H}_{\uparrow}$ enforces stability with respect to other interactions $\delta H_{I}$ given by small changes of the potential $V_I(x,y)$ or small interactions with other degrees of freedom denoted by $z$. Indeed, in the Heisenberg picture 
\begin{equation}
\label{eq:Ehrenfest}
\frac{d}{dt}\langle \hat{H}_{\uparrow}(t)\rangle =i\langle [\delta \hat{H}_{I},\hat{H}_{\uparrow}(t)]\rangle\,,
\end{equation}
so that for sufficiently small $\delta \hat{H}_{I}$ the expectation value of $\hat{H}_{\uparrow}(t)$ in $\Psi_{0}\left(x,y\right)$ may not reach the next level $E_{1}^{\uparrow}$ even after parametrically long times. The same is applicable for fluctuations. Moreover, for a finite duration of interaction the change of the expectation value can be parametrically smaller than the gap $E_{1}^{\uparrow}-E_{0}^{\uparrow}$, making the probability of the transition negligible. For illustration of this stability argument on example of $\delta \hat{H}_{I}=\alpha x^{2}y^{2}$, see Fig.~(\ref{fig:confinement_pert}). Finally, it is important to mention that all eigenstates of $\hat{H}$ possibly located in a small energy interval around vacuum are very highly excited states of $\hat{H}_{\uparrow}$ and $\hat{H}_{\downarrow}$ with very large $n$ and $m$. According to Eq.~\eqref{eq:Ehrenfest}, short-time perturbations never allow for transitions to these highly excited states from vacuum or other low $(n,m)$-states. On the other hand, long-lived perturbations require parametrically long times for such transitions.  

\begin{figure}[ht]
    \centering
    \resizebox{\columnwidth}{!}{
    \includegraphics{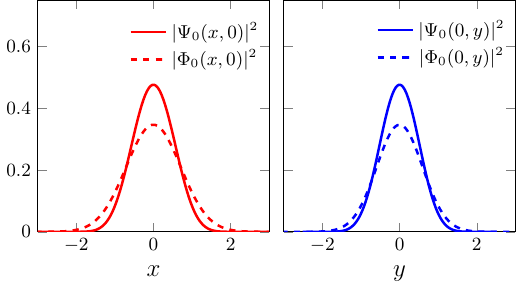}
    }
    \caption{Side-view comparisons of probability densities for $\Psi_0$ (solid) and $\Phi_0$ (dashed). Thus, for system \eqref{eq:system}, the ground state of the interacting ghost, $\Psi_0$, is more confining than the ground state $\Phi_0$ of two corresponding decoupled oscillators $x$ and $y$.}
    \label{fig:overlay}
\end{figure}

\begin{figure}[ht]
    \centering
    \resizebox{\columnwidth}{!}{
    \includegraphics{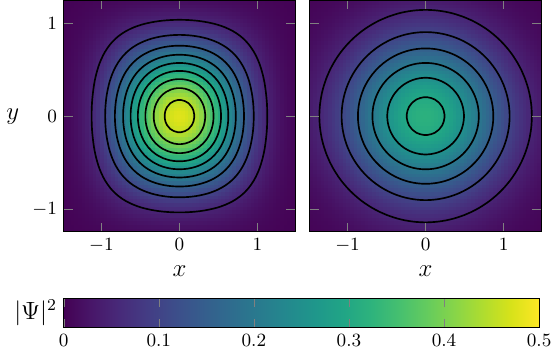}
    }
    \caption{Top-down view of ground state probability densities: interacting, $|\Psi_{0}|^2$, (left) and decoupled, $|\Phi_{0}|^2$, (right), both with imposed contours. This is another way to see that for system \eqref{eq:system} the interacting vacuum with a ghost $\Psi_0$ is more confining than the free vacuum $\Phi_0$.}
    \label{fig:contours}
\end{figure}

\begin{figure}[ht!]
    \centering
    \resizebox{\columnwidth}{!}{
    \includegraphics{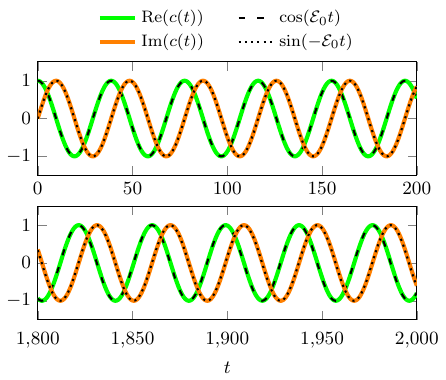}
    }
    \caption{
    Overlap ${c(t)\equiv\braket{\Psi_0(0)}{\Psi_0(t)}}$ of the numerically computed ground state $\Psi_0$ and its time evolution ${\Psi_0(t)}$, with energy ${\mathcal{E}_0\approx-0.162}$. For comparison, we include the expected dashed curves ${\cos(\mathcal{E}_0t)}$ and ${\sin(-\mathcal{E}_0t)}$.}\label{fig:sine-cosine}
\end{figure}

\begin{figure}[ht!]
    \centering
    \resizebox{\columnwidth}{!}{
    \includegraphics{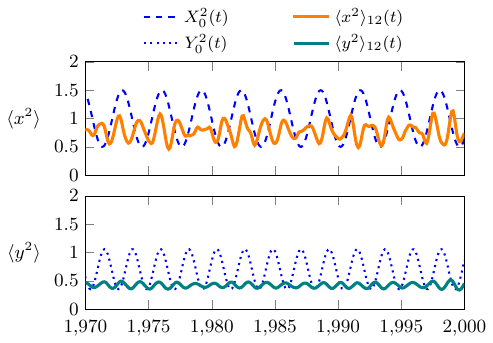}
    }
    \caption{The expectation values ${\langle x^2\rangle_{12}(t)}$, ${\langle y^2\rangle_{12}(t)}$, and decoupled trajectories ${X^2_0(t)}$, ${Y^2_0(t)}$. We omit plotting ${b=6}$, as the difference in graphs is visually indistinguishable: ${\underset{t\in[0,T]}{\max}{\left|\frac{\langle x^2\rangle_{12}-\langle x^2\rangle_{6}}{\langle x^2\rangle_{12}}\right|}\sim 10^{-3}}$ and ${\underset{t\in[0,T]}{\max}{\left|\frac{\langle y^2\rangle_{12}-\langle y^2\rangle_{6}}{\langle y^2\rangle_{12}}\right|}\sim 10^{-3}}$.}\label{fig:confinement}
\end{figure}

\begin{figure}[ht!]
    \centering
    \resizebox{\columnwidth}{!}{
    \includegraphics{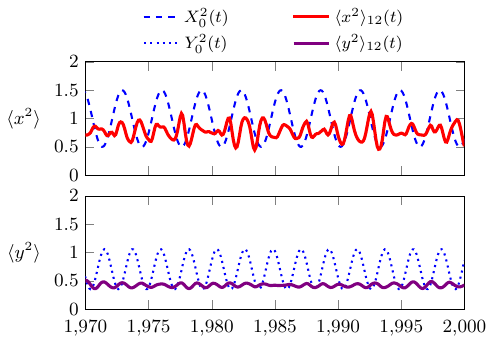}
    }
    \caption{Same as Fig.~\ref{fig:confinement} but with an additional, interaction term $\delta \hat{H}_{I}=\alpha x^{2}y^{2}$ for coupling constant $\alpha=0.01$.}\label{fig:confinement_pert}
\end{figure}

{ \textbf{\textit{Numerical Investigation.}} }
All numerical simulations are run with ${\omega=1.414213}$ and non-negligible couplings ${(\gamma,c)=(0.1,5)}$. To find the ground state $\Psi_{0}(x,y)$ and the corresponding eigenvalue, $\mathcal{E}_{0}$, of the Hamiltonian $\hat{H}$, we implemented a pseudo-spectral method to solve the time-independent Schr\"odinger equation for the integrals of motion ${\hat{H}_{\uparrow\downarrow}}$, see \cite{Boyd:2001}. \Cref{fig:overlay} shows side-view plots of the numerically computed ground state $\Psi_{0}(x,y)$, and compares it to $\Phi_{0}(x,y)$ given by Eq.~\eqref{eq:nonint_gs}. \Cref{fig:contours} gives a top-down view with imposed contours to illustrate their difference in shapes. Overall, we see that $\Psi_0(x,y)$ enjoys the same $\mathbb{Z}_2$-symmetries as $V_I(x,y)$, and that its more squared-off shape reflects the shape of the potential. Surprisingly, the interacting vacuum $\Psi_0(x,y)$ involving the ghost is more confining than the free vacuum of two oscillators $\Phi_0(x,y)$. It is worth noting that, numerically obtained ${\mathcal{E}_0\approx-0.162\neq \frac{1}{2}\left(1-\omega\right)\approx-0.207}$, but, as expected, $\mathcal{O}\left(\gamma,c^{-1}\right)$ close to the non-interacting value. 

A second-order split-operator method was used for solving the time-dependent Schr\"odinger equation \eqref{eq:Sch}, see \cite{Hairer:2006}. To begin with, we assess the validity of our stationary numerical solution by evolving the ground state $\Psi_{0}$ in time. Given ${\Psi_0(t)\equiv e^{-i\hat{H}t}\Psi_0}(0)$, we define the \textit{overlap} ${c(t)\equiv \braket{\Psi_0(0)}{\Psi_0(t)}}$. Since $\Psi_0$ is an eigenstate of $\hat{H}$ with energy ${\mathcal{E}_0}$, we expect ${c(t)\approx e^{-i\mathcal{E}_0t}}$ up to numerical errors. \Cref{fig:sine-cosine} tracks the real and imaginary parts of $c(t)$ over approximately $50$ periods of oscillation, $2\pi/|\mathcal{E}_0|$, and compares $c(t)$ with the expected trivial behavior; the two are visually indistinguishable\footnote{Note that ${\mathcal{E}_0}$ is negative for this particular choice of parameters, which introduces an extra negative sign in the imaginary part.}. 
To illustrate bounding of the motion in \eqref{eq:bound}  we consider a Gaussian displaced to ${(x_0,y_0)}$ with initial momentum $(p_{0x},p_{0y})$, as an initial state:
\begin{equation}
\label{eq:init_cond}
    \Psi_{\text{G}}(x,y)=\left(\frac{\omega}{\pi^{2}}\right)^{\frac{1}{4}}e^{-\frac{1}{2}(x-x_{0})^{2}-\frac{1}{2}\omega(y-y_{0})^{2}+i\left(xp_{0x}+yp_{0y}\right)}\,.    
\end{equation}
We choose ${x_0=1}$, ${y_0=0}$, and ${p_{0x}=0}$, ${p_{0y}=1}$, and solve \eqref{eq:Sch} numerically\footnote{For animations of the evolution of probability density $|\Psi_{\text{G}}(x,y,t)|^2$ see, \href{https://doi.org/10.5281/zenodo.19710447}{https://doi.org/10.5281/zenodo.19710447}. }  until $T=2000$ with time step $\Delta t=0.01$\footnote{For reference, $T_{\text{G}}={2\pi}{/|\langle E\rangle|}\approx{2\pi}{/2.938}\approx 2.138$, is a characteristic time-scale for $\Psi_{\text{G}}(t)$. It would be the period of oscillations of $\Psi_{\text{G}}(t)$, if it were actually an eigenstate of $\hat{H}$. So ${T=2000\approx 10^3\cdot T_{\text{G}}}$.}. In the decoupled case the state $\Psi_{\text{G}}(x,y,t)$ would be a coherent state. Then, the corresponding expectation values ${\langle x\rangle}\propto\cos(t)$ and ${\langle y\rangle}\propto\sin(-t)$ would follow the classical trajectories, and we have that
\begin{equation}
    \langle x^2\rangle= \frac{1}{2}+\cos^2(t)\,,\qquad
    \langle y^2\rangle=\frac{1}{2\omega}+\frac{1}{\omega}\sin^2(t)\,,
\end{equation}
where, as usual for coherent states, the first terms correspond to vacuum quantum fluctuations. We denote these decoupled "trajectories" by ${X^2_0(t)}$ and ${Y^2_0(t)}$. 

Fig.~\ref{fig:confinement} shows ${\langle x^2\rangle_b}$ and ${\langle y^2\rangle_b}$ obtained from a numerical solution of the time-dependent Schr\"odinger equation with initial state \eqref{eq:init_cond} in the square $(-b,b)\times (-b,b)$ with $b=12$, as well as ${X^2_0(t)}$ and ${Y^2_0(t)}$, during the last $30$ units of time evolution. We do not plot ${b=6}$, as it is visually indistinguishable from ${b=12}$. This demonstrates that ${\langle x^2\rangle}$ and ${\langle y^2\rangle}$ are independent of $b$. Moreover, since $X^2_0$ and $Y^2_0$ bound ${\langle x^2\rangle}$ and ${\langle y^2\rangle}$ from above, we conclude that the interaction potential $V_I(x,y)$ (or more intuitively the resulting positive definite integral of motion ${\hat{H}_\uparrow}$) is indeed \textit{more} confining than two decoupled harmonic oscillator potentials. 

To illustrate discussion around \eqref{eq:Ehrenfest}, we numerically solved the time-dependent Schr\"odinger equation with initial state \eqref{eq:init_cond} for additional interaction $\delta \hat{H}_{I}={\alpha x^2y^2}$. For $b=12$, we plotted the resulting ${\langle x^2\rangle_{12}(t)}$ and ${\langle y^2\rangle_{12}(t)}$ in Fig.~\ref{fig:confinement_pert} which clearly demonstrates that the motion is still bounded and barely distinguishable from the one without $\delta \hat{H}_{I}$, despite the perturbative and nonintegrable modification of interaction. 


{ \textbf{\textit{Conclusions.}} } We have canonically quantized a globally stable classical system introduced in \cite{Deffayet:2023wdg} and consisting out of a ghost oscillator polynomially coupled to a usual harmonic oscillator, both with self-interactions. The classical motion of this system is confined for all initial data - proceeds in a finite region of phase space for all times. It is the standard wisdom that systems with finite motion have discrete spectra of Hamiltonians upon quantization, see e.g. \cite{Simon:1983jy}. In this work we confirmed this intuition for systems involving interacting ghosts and showed that the self-adjoint Hamiltonian has a pure point spectrum even though it is unbounded from above and below.  
The key to prove these features is the existence of the self-adjoint and positive definite quantum integral of motion $\hat{H}_{\uparrow}$, which has a purely discrete spectrum and replaces the energy in confining the motion. In particular, the lowest eigenstate of this operator defines the unique and positive ground state $\Psi_{0}\left(x,y\right)$. 

Crucially, the energy spectrum \eqref{eq:E_spectrum} is not continuous. Yet, a continuous spectrum is what one would expect for the unstable or unbounded classical motion usually associated with interacting ghosts, see e.g.~\cite{Pais:1950za}. However, whether the spectrum of $\hat{H}$ is \textit{dense}, as is generically the case for two decoupled harmonic oscillators, with one being a ghost (see e.g. \cite{Smilga:2008pr,Smilga:2017arl,Damour:2021fva}), or contains accumulation points, is carefully discussed in a separate publication \cite{Deffayet:toAppear}.  

We numerically computed the ground state and surprisingly found that the interacting vacuum state is more confining than the decoupled ones, see Fig.~\ref{fig:overlay}. Moreover, we solved the time-dependent Schr\"odinger equation to assess validity and numerical stability of the ground state, see Fig.~\ref{fig:sine-cosine}, and to illustrate physical confinement \eqref{eq:bound} of canonical variables, see Fig.~\ref{fig:confinement}. Finally, we argued that the discrete spectrum of $\hat{H}_{\uparrow}$ 
enforces stability with respect to small additional interactions, either of the present degrees of freedom or of these degrees of freedom with external ones, see Fig.~\ref{fig:confinement_pert}. Moreover, one can argue that the presence of the gap can make the stability even stronger than in the classical case. 
Thus, in this paper, we established that the quantum interacting ghosts have the potential to model meaningful physical systems.  

{ \textbf{\textit{Note added.}} }
In the last moments before submission we became aware of work 
\cite{Ewasiuk:toAppear} where similar conclusions are obtained for the system from \cite{Deffayet:2021nnt}. This system differs from the one studied here, as there the interaction between ghost and normal oscillator is non-polynomial, is bounded everywhere in configuration space and vanishes in the limit of large coordinates.  

{ \textbf{\textit{Acknowledgments.}} }
We are grateful to Christopher Ewasiuk and Stefano Profumo for sharing their results \cite{Ewasiuk:toAppear} prior to publication. The work of S.~M. was supported in part by Japan Society for the Promotion of Science (JSPS) Grants-in-Aid for Scientific Research No.~24K07017 and the World Premier International Research Center Initiative (WPI), MEXT, Japan. 
The work of A.~V. and of A.~F.~J. was supported by European Structural and Investment Funds and the Czech Ministry of Education, Youth and Sports (Project FORTE CZ.02.01.01/00/22 008/0004632).

\bibliographystyle{utphys}
\bibliography{bibliography}  
\end{document}